\pdfoutput=1
\documentclass[]{article}  
\usepackage{url,float}
\usepackage{graphicx}
\usepackage{amsmath}
\usepackage{amsfonts}
\usepackage{amssymb}
\usepackage{latexsym}
\usepackage{mathtools}
\usepackage{color}
\usepackage[boxruled]{algorithm2e}

\newcommand{\hide}[1]{}

\newcommand{\ABox}{
\raisebox{3pt}{\framebox[6pt]{\rule{6pt}{0pt}}}
}
\newenvironment{proof}{{\bf Proof:}}{\hfill\ABox}

\newtheorem{thm}{{\bf Theorem}}

\newtheorem{lem}{Lemma}

\newcommand{\lemlab}[1]{\label{lemma:#1}}
\newcommand{\thmlab}[1]{\label{thm:#1}}

\newcommand{\figlab}[1]{\label{fig:#1}}
\newcommand{\seclab}[1]{\label{sec:#1}}

\newcommand{\thmref}[1]{\ref{thm:#1}}

\newcommand{\secref}[1]{\ref{sec:#1}}
\newcommand{\figref}[1]{\ref{fig:#1}}

\def\a{{\alpha}}

\def\o{{\omega}}

\def\d{{\delta}}

\newcommand{\squeezelist}{\setlength{\itemsep}{0pt}}

\usepackage{enumitem}

\title{%
Every Combinatorial Polyhedron\\
Can 
Unfold with Overlap
} 

\author{%
Joseph O'Rourke%
  \thanks{Departments of Computer Science and of Mathematics, Smith College, Northampton, MA
      01063, USA.
      \protect\url{jorourke@smith.edu}.}
}

\date{\today}

\begin{document}
\maketitle

\begin{abstract}
Ghomi proved that every convex polyhedron could be stretched via an affine transformation
so that it has an edge-unfolding to a net~\cite{g-aucp-14}.
A \emph{net} is a simple planar polygon; in particular, it does not self-overlap.
One can view his result as establishing that every combinatorial polyhedron has a
metric realization that allows unfolding to a net.

Joseph Malkevitch asked if the reverse holds (in some sense of ``reverse"):
Is there a combinatorial polyhedron $\mathcal{P}$ such that, for every metric realization
$P$ in $\mathbb{R}^3$, and for every
spanning cut-tree $T$, $P$ cut by $T$ unfolds to a net?
In this note we prove the answer is \textsc{no}:
every combinatorial polyhedron has a realization and a cut-tree that unfolds 
the polyhedron with overlap.
\end{abstract}

\section{Introduction}
\seclab{Introduction}
Joseph Malkevitch asked\footnote{Personal communication, Dec. 2022} 
whether there is a combinatorial $\mathcal{P}$ type of a convex polyhedron $P$ in $\mathbb{R}^3$ 
whose every edge-unfolding results in a net. One could imagine, to use his example, that every
realization of a combinatorial cube unfolds without overlap for each of its $384$ spanning cut-trees~\cite{tuffley2011counting}.%
\footnote{
Burnside's Lemma can show that these $384$ trees lead to 
$11$ incongruent unfoldings of the cube~\cite{goldstone2019unfoldings}.}
The purpose of this note is to prove this is, alas, not true:
every combinatorial type can be realized and edge-unfolded to overlap:
Theorem~\thmref{main} (Section~\secref{Theorem}).
For an overlapping unfolding of a combinatorial cube, see ahead to Fig.~\figref{CubeUnf}.

An implication of Theorem~\thmref{main}, together with~\cite{g-aucp-14}, is that the resolution of D\"urer's Problem~\cite{o-dp-13}
must focus on the geometry rather than the combinatorial structure of convex polyhedra.

\section{Proof Outline}
\seclab{ProofOutline}
We describe the overall proof plan in the form of a multi-step algorithm. 
We will illustrate the steps with an icosahedron
before providing details.

\begin{algorithm*}
\DontPrintSemicolon 
\KwIn{A $3$-connected planar graph $G$.}
\KwOut{Polyhedron $P$ realizing $G$ and a cut-tree $T$ that unfolds $P$ with overlap.}
(1)~Select outer face $B$ as base.\;
(2)~Embed $B$ as a convex polygon in the plane.\;
(3)~Apply Tutte's theorem to calculate an equilibrium stress for $G$.\;
(4)~Apply Maxwell-Cremona lifting to $P$.\;
(5)~Identify special triangle $\triangle$.\;
(6)~Scale $P$ horizontally (if necessary).\;
(7)~Scale $P$ vertically (if necessary).\;
(8)~Form cut-tree $T$, including `\texttt{Z}' around $\triangle$.\;
(9)~Unfold $P \setminus T$.\;
(10)~$\to$ Overlap.
\caption{Realizing $G$ to unfold with overlap.}
\label{Algo}
\end{algorithm*}

We are given a $3$-connected planar graph $G$, which constitutes the combinatorial type
of a convex polyhedron. 
By Steinitz's theorem, we know $G$ is the $1$-skeleton of a convex polyhedron.
Initially assume $G$ is triangulated;
this assumption will be removed in Section~\secref{NonTri}.

\begin{enumerate}[label={(\arabic*)}]
\item Select outer face $B$ as base.
Initially, any face suffices. Later we will coordinate the choice of $B$ with the choice of the
special triangle $\triangle$.
\item Embed $B$ as a convex polygon in the plane.
Select coordinates for the vertices of $B$, which then pin $B$ to the plane. $B$ must be
convex, but otherwise its shape is arbitrary.
\item Apply Tutte's theorem~\cite{t-hdg-63}
to calculate an equilibrium stress---positive 
weights on each edge---that, when interpreted as forces, induce an equilibrium 
(sum to zero) at every vertex.
This provides explicit coordinates for all vertices interior to $B$.
The result is a Schlegel diagram, with all interior faces convex regions.
Fig.~\figref{IcosaSchlegel} illustrates this for the icosahedron.\footnote{
Here the drawing is approximate, because I did not explicitly calculate the equilibrium stresses.}
\begin{figure}[htbp]
\centering
\includegraphics[width=0.75\linewidth]{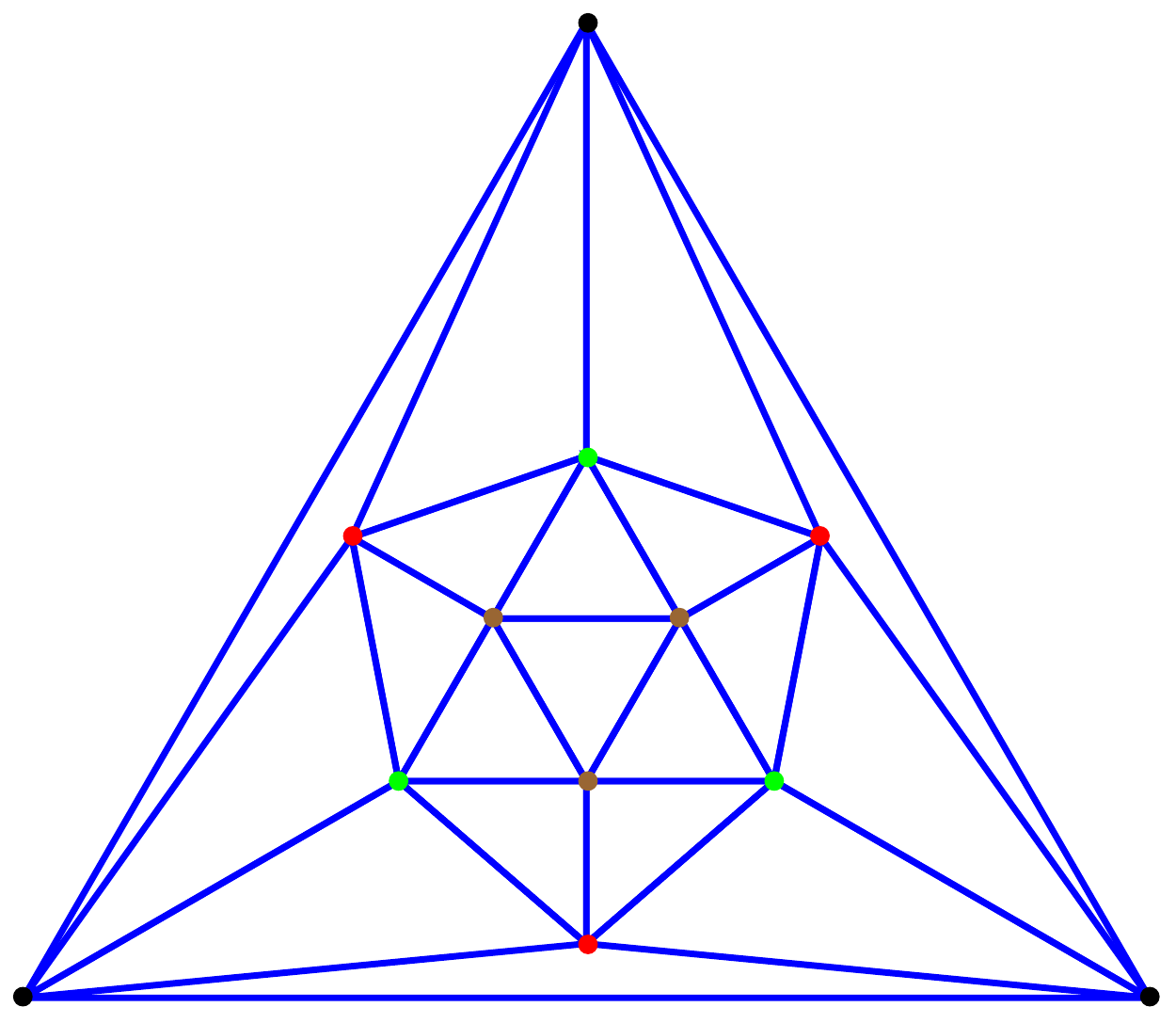}
\caption{Icosahedron Schlegel diagram.}
\figlab{IcosaSchlegel}
\end{figure}
%
\item Apply Maxwell-Cremona lifting to $P$.
The Maxwell-Cremona theorem says that any straight-line planar drawing with an equilibrium stress has a polyhedral lifting
via a ``reciprocal diagram."
The details are not needed here;\footnote{
A good resource on this topic is~\cite{richter2006realization}.}
we only need the resulting lifted polyhedron.
An example from~\cite{schulz2008lifting} shows the lifting of a Schlegel diagram of
the dodecahedron: Fig.~\figref{AndreSchultzDodeca}.
A lifting of the vertices of the icosahedron in Fig.~\figref{IcosaSchlegel}
is shown in Fig.~\figref{MCLift}.\footnote{This is again an approximation as I did not
calculate the reciprocal diagram.}
\begin{figure}[htbp]
\centering
\includegraphics[width=0.75\linewidth]{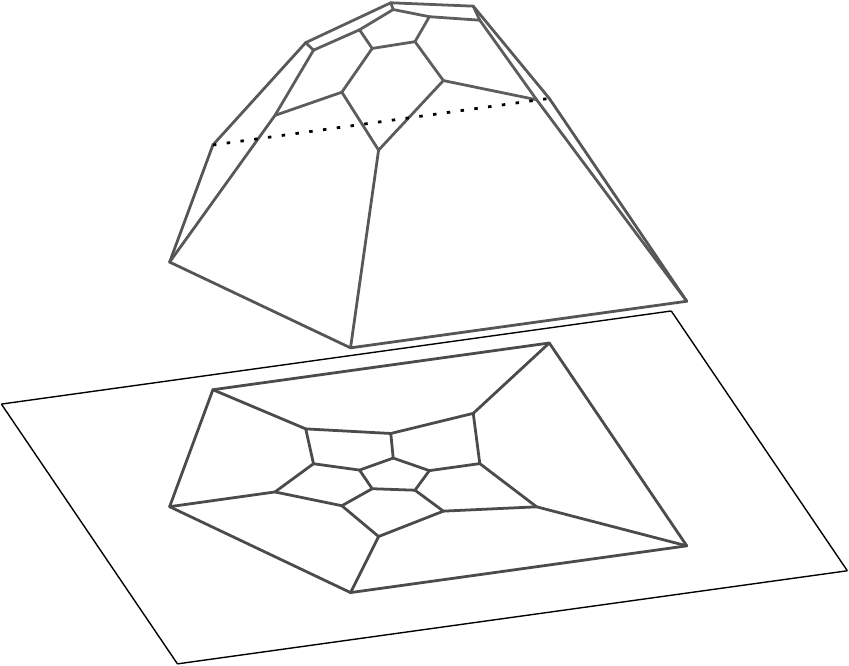}
\caption{Maxwell-Cremona lifting to a dodecahedral diagram.
\protect\cite{schulz2008lifting}, by permission of author.}
\figlab{AndreSchultzDodeca}
\end{figure}

\begin{figure}[htbp]
\centering
\includegraphics[width=0.75\linewidth]{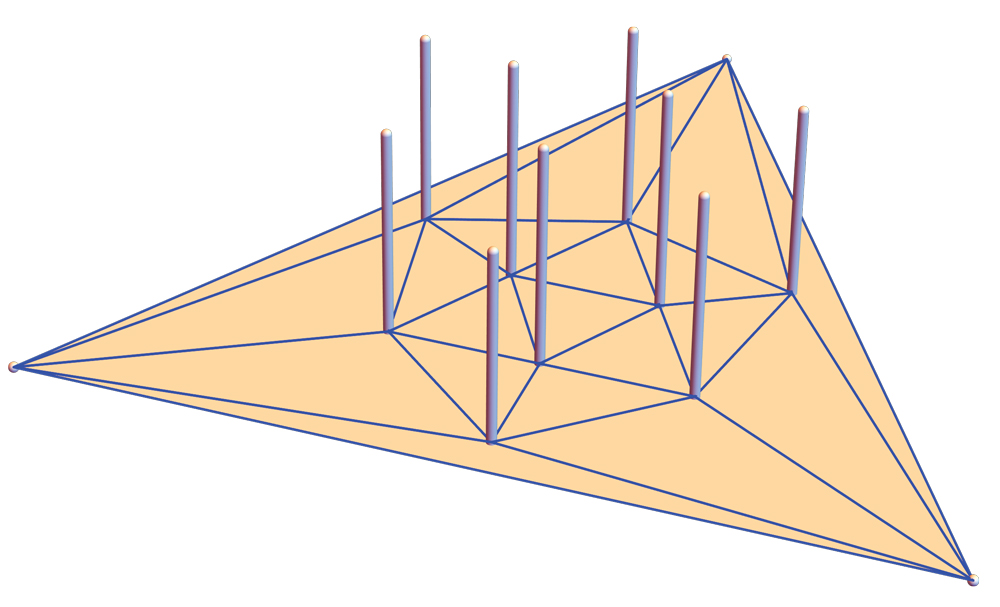}
\caption{Lifting the vertices of the icosahedron Schlegel diagram in~Fig.~\protect\figref{IcosaSchlegel}.}
\figlab{MCLift}
\end{figure}

\item Identify special triangle $\triangle$.
This special triangle must satisfy several conditions, which we detail later (Section~\secref{ConditionsTriangle}).
For now, we select $\triangle = a_1 a_2 a_3 = 6,8,5$ in Fig.~\figref{IcosaUnf_1}.
\begin{figure}[htbp]
\centering
\includegraphics[width=0.6\textheight]{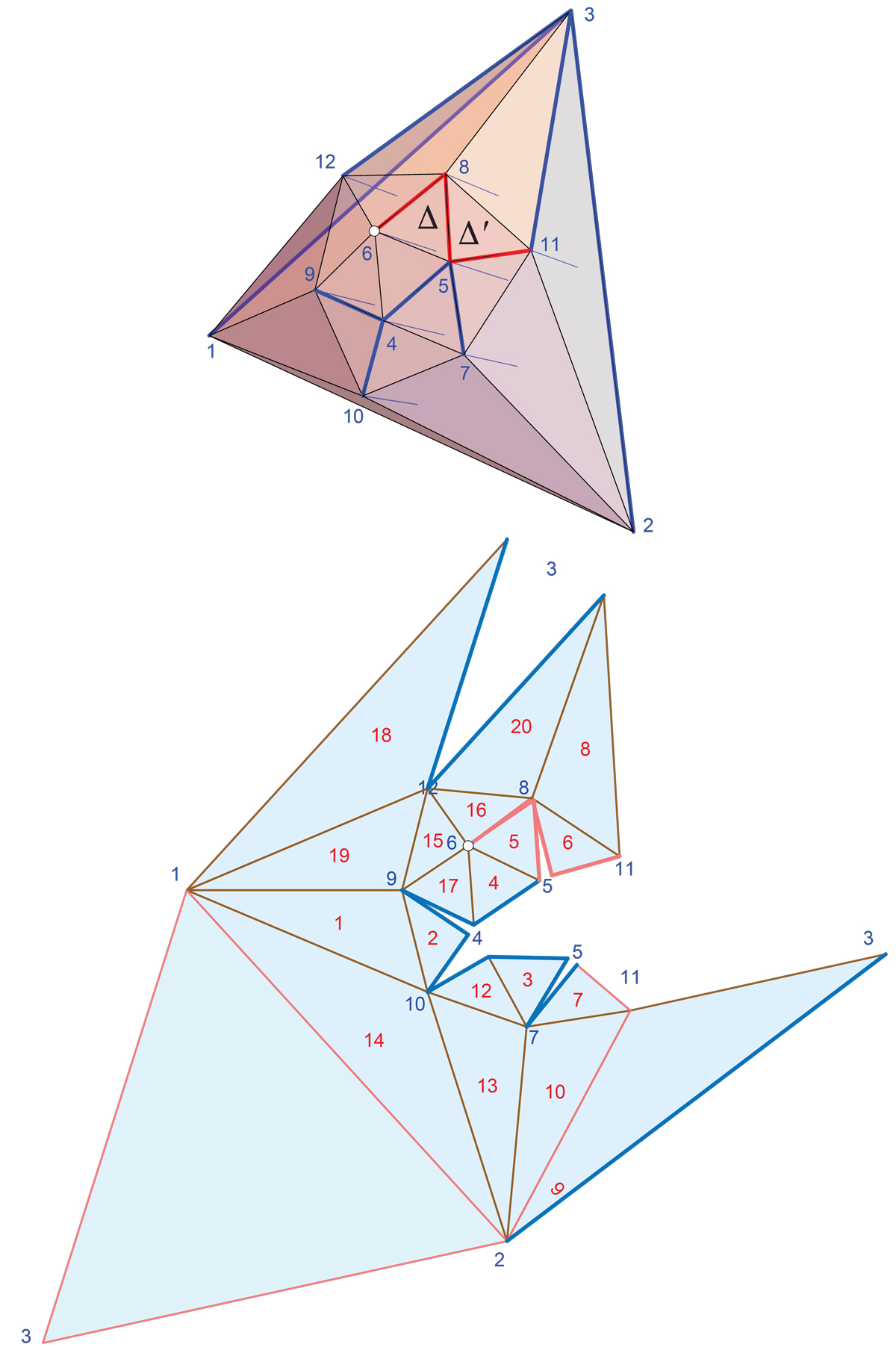}
\caption{Red: face numbers; blue: vertex indices.
$\triangle=5$, $\triangle'=6$.
\texttt{Z}-portion of spanning tree $T$ red; remainder blue.}
\figlab{IcosaUnf_1}
\end{figure}

\item Scale $P$ horizontally (if necessary).
Not needed in icosahedron example.
\item Scale $P$ vertically (if necessary).
Not needed in icosahedron example.
\item Form cut-tree $T$, including a `\texttt{Z}'-path around $\triangle$.
We think of $a_1$ as the root of the spanning tree, which includes 
the \texttt{Z}-shaped (red) path $a_1 a_2 a_3 a_4$ around $\triangle$
and the adjacent triangle $\triangle'$ sharing edge $a_2 a_3$.
In Fig.~\figref{IcosaUnf_1}, the \texttt{Z} vertex indices are $6,8,5,11$.
The remainder of $T$ is completed arbitrarily.
\item Unfold $P \setminus T$.
\item
Finally, the conditions on $\triangle$ ensure that cutting $T$ unfolds
$P$ with overlap along the $a_2 a_3$ edge.
See Fig.~\figref{IcosaUnf_2}.
\begin{figure}[htbp]
\centering
\includegraphics[width=1.0\linewidth]{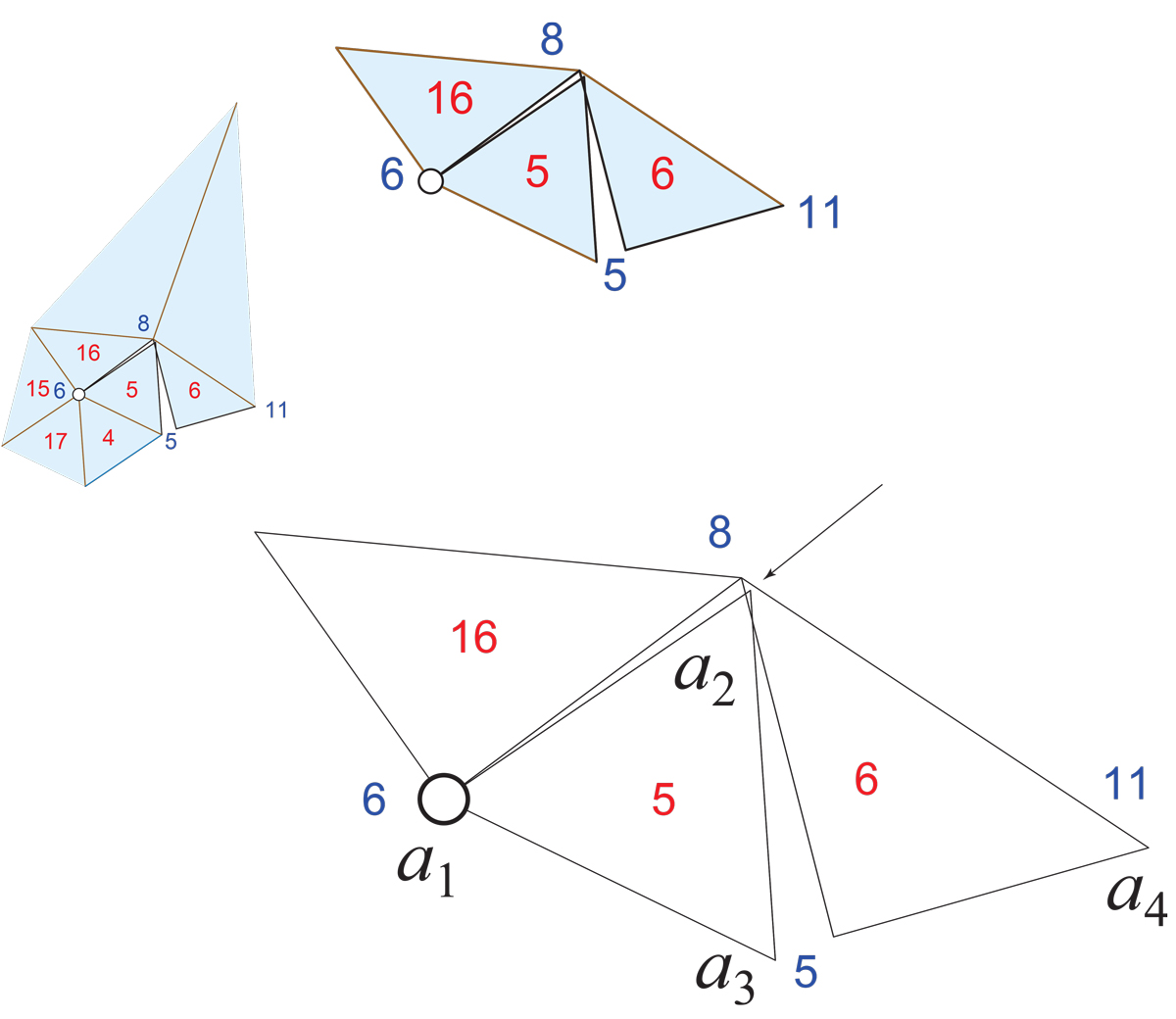}
\caption{Close-up views of overlap.}
\figlab{IcosaUnf_2}
\end{figure}
\end{enumerate}

\clearpage
\section{Conditions on $\triangle$}
\seclab{ConditionsTriangle}
We continue to focus on triangulated polyhedra.
In order to guarantee overlap, the special triangle $\triangle = a_1 a_2 a_3$ should satisfy
several conditions:
\begin{enumerate}
\item The angle at $a_2$ in $\triangle$ must be $\le \pi/3 = 60^\circ$,
and the edge $a_2 a_3$ at least as long as $a_1 a_2$.
\item The spanning cut-tree $T$ must contain the \texttt{Z} as previously explained.
In addition, no other edge of $T$ is incident to either $a_1$ or $a_2$.
In particular, edge $a_1 a_3$ is not cut, so the triangle $\triangle$ rotates as a unit about $a_1$.
\item The curvatures at $a_1$ and $a_2$ must be small. 
(The \emph{curvature} or ``angle gap" at a vertex is $2\pi$ minus the sum of the incident face angles.)
We show below that $< 20^\circ$ suffices.
\item $\triangle$ should be disjoint from the base $B$: $\triangle$ and $B$ share no vertices.
\end{enumerate}
This 4th condition might be impossible to satisfy,
in which case an additional argument is needed (Section~\secref{Disjoint}).
For now we concentrate on the first three conditions.

$\triangle$ is chosen to be the triangle disjoint from $B$ with the smallest angle $\a$.
Clearly $\a \le \pi/3 = 60^\circ$.
Let $\triangle=a_1 a_2 a_3$ with $a_2$ the smallest angle.
Chose the labels so that $|a_1 a_2| \le |a_2 a_3|$.
It will be easy to see that $\triangle$ an equilateral triangle 
is the ``worst case" in that smaller $\a$ leads to deeper overlap,
and $|a_1 a_2| = |a_2 a_3|$ suffices for overlap.
So we will assume $\triangle$ is an equilateral triangle.

Next, we address the requirement for small curvatures,
when the second condition is satisfied: 
no other edge of $T$ is incident to either $a_1$ or $a_2$.
Let $\o$ be the curvature at both $a_1$ and $a_2$.
Then an elementary calculation shows that
$\o=\frac{1}{9} \pi = 20^\circ$
would just barely avoid overlap:
see Fig.~\figref{Omega2010}.
\begin{figure}[htbp]
\centering
\includegraphics[width=1.0\linewidth]{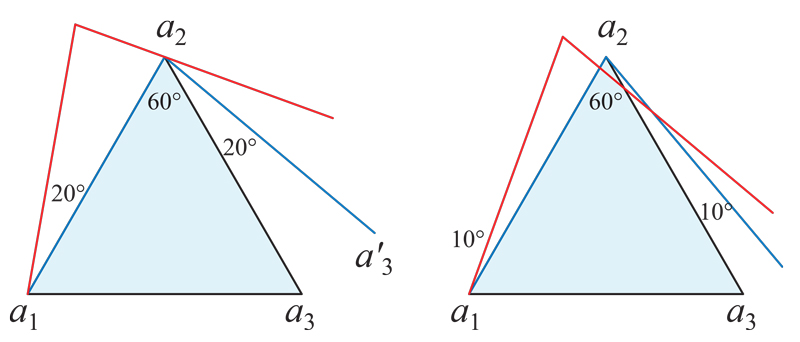}
\caption{Left: $\o= 20^\circ$ avoids overlap.
Right: $\o= 10^\circ$ overlaps.}
\figlab{Omega2010}
\end{figure}

One can view the flattening of $a_1$ and $a_2$ when cut
as first turning the edge $a_2 a_3$ by $\o$ about $a_2$,
and then rotating the rigid path $a_1 a_2 a'_3$ about $a_1$ by $\o$.
For any $\o$ strictly less than $20^\circ$, overlap occurs along the $a_2 a_3$ edge.
The basic reason this ``works" to create overlap is that
the cut-path around $\triangle$ is not \emph{radially monotone},
a concept introduced in~\cite{o-ucprm-16} and used in~\cite{o-eunfcc-17}
and~\cite{radons2021edge} to avoid overlap.

In the unfolded icosahedron in Fig.~\figref{IcosaUnf_1}, 
the angle at $a_2$ is $59^\circ$, and
the curvatures $\o_1,\o_2$ at $a_1,a_2$ are 
$2.4^\circ$ and $8.1^\circ$ respectively.

If the two curvatures are not less than $20^\circ$, then we
scale $P$ vertically, orthogonal to base $B$ (step (7) of Algorithm~\ref{Algo}.
As illustrated in Fig.~\figref{Dihedrals3}, this flattens dihedral angles
and reduces vertex curvatures at all but the vertices of base $B$,
which increase to compensate the Guass-Bonnet sum of $4\pi$.
Clearly we can reduce curvatures as much as desired.
\begin{figure}[htbp]
\centering
\includegraphics[width=1.0\linewidth]{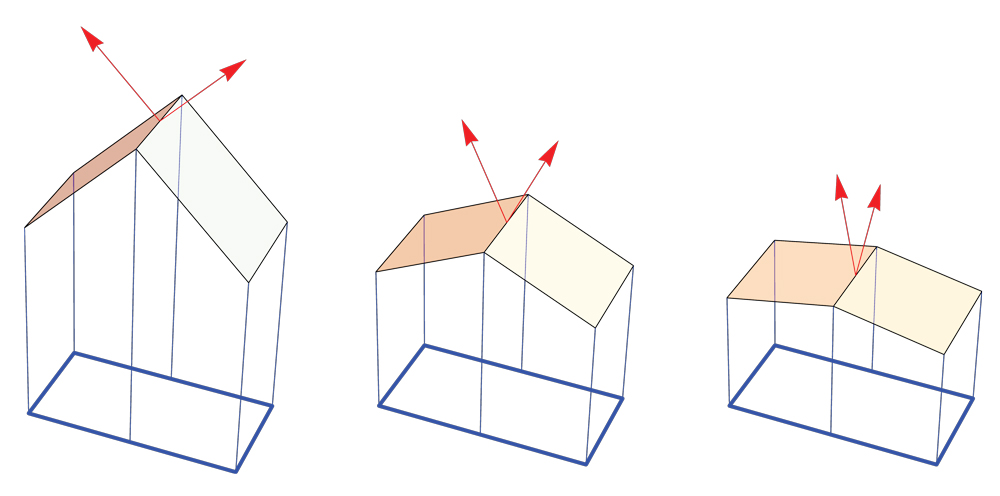}
\caption{Dihedral angle $\d$ flattens as $z$-heights scaled:
$(1, \frac{1}{2},\frac{1}{5} ) \to (90^\circ, 125^\circ, 160^\circ)$.}
\figlab{Dihedrals3}
\end{figure}

\subsection{Non-Triangulated Polyhedra}
\seclab{NonTri}
If $G$ and therefore $P$ contains non-triangular faces,
then we employ
step (6) of Algorithm~\ref{Algo}: Scale $P$ horizontally, parallel to the $xy$ plane containing $B$.
For example, in the dodecahedron example (Fig.~\figref{AndreSchultzDodeca}), no face has an angle 
$\a \le \pi/3$. 
But by horizontal scaling, we can sharpen any selected face angle,
as illustrated in Fig.~\figref{PentScaled}.
Then we can identify $\triangle$ within that face, and proceed just as in a
triangulated polyhedron.
\begin{figure}[htbp]
\centering
\includegraphics[width=0.75\linewidth]{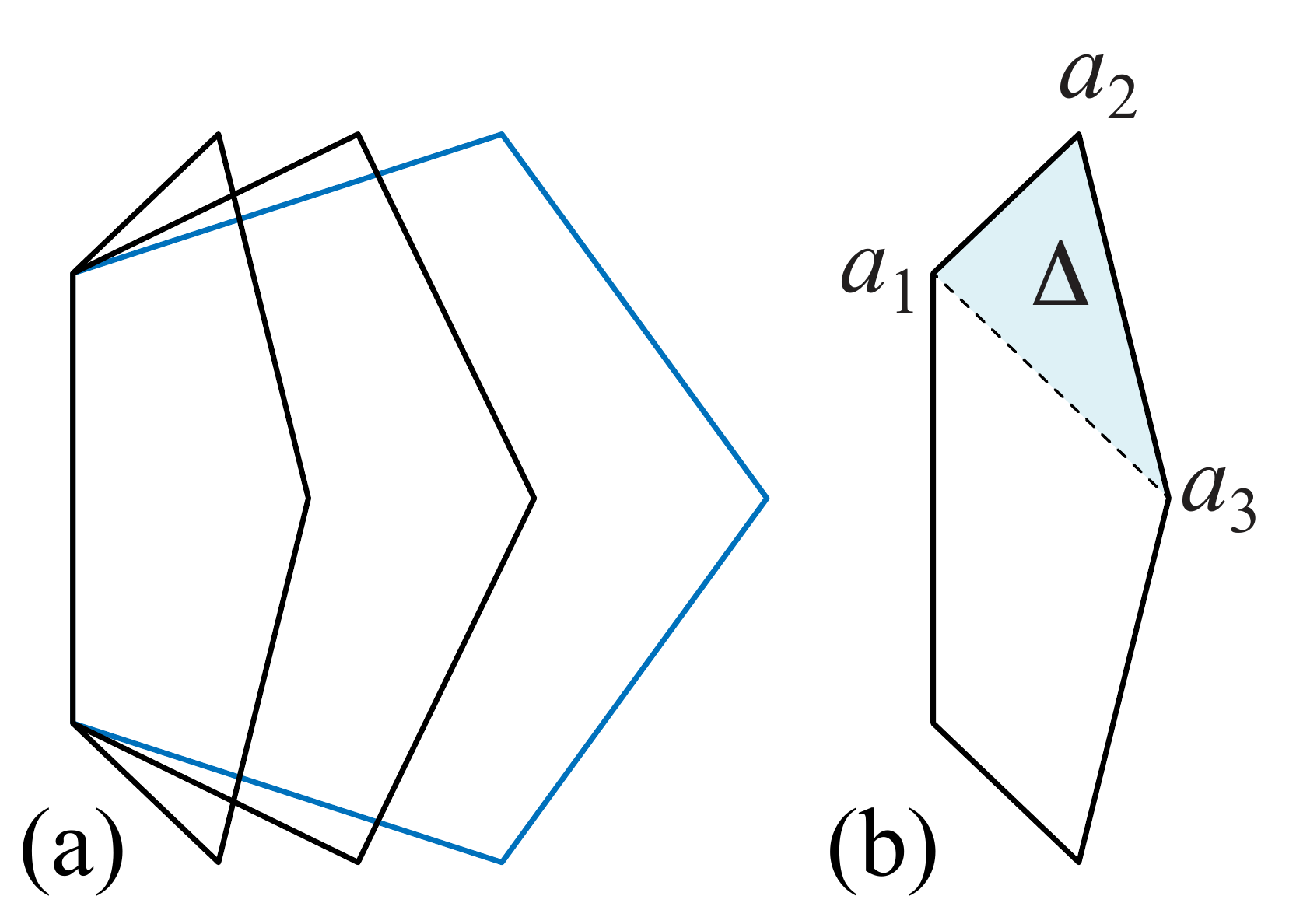}
\caption{(a)~Regular pentagon scaled $\frac{2}{3}$ and $\frac{1}{3}$ horizontally.
(b)~A triangle with one angle $60^\circ$.}
\figlab{PentScaled}
\end{figure}

\clearpage

\section{No Pair of Disjoint Faces}
\seclab{Disjoint}
Finally we focus on the 4th condition that
$\triangle$ should be disjoint from the base $B$.
If $G$ contains any two disjoint faces, triangles or $k$-gon faces with $k > 3$,
we select one as $B$ and the other to yield $\triangle$.
So what remains is those $G$ with no pair of disjoint faces.

For example, a pyramid---$B$ plus one vertex $a$ (the apex) above $B$---has no pair
of disjoint faces. However, note that a pyramid has pairs of faces that share one vertex but not two vertices.
It turns out that this suffices to achieve the same structure of overlap.
Fig.~\figref{PlatformNet_SmallF} illustrates why.
Here $B$ is a triangle $b_1 b_2 a_3$ and we select $\triangle= a_1 a_2 a_3$.
The small-curvature requirement holds just for $a_1, a_2$---the start of the \texttt{Z}---the 
curvature at $a_3$ could be large ($117^\circ$ in this example) 
but does not play a role, as the unfolding illustrates.
Therefore, if $G$ has no pair of disjoint faces, but does have a pair of faces
that share a single vertex, we proceed just in Algorithm~\ref{Algo},
suitably modified.
\begin{figure}[htbp]
\centering
\includegraphics[width=1.0\linewidth]{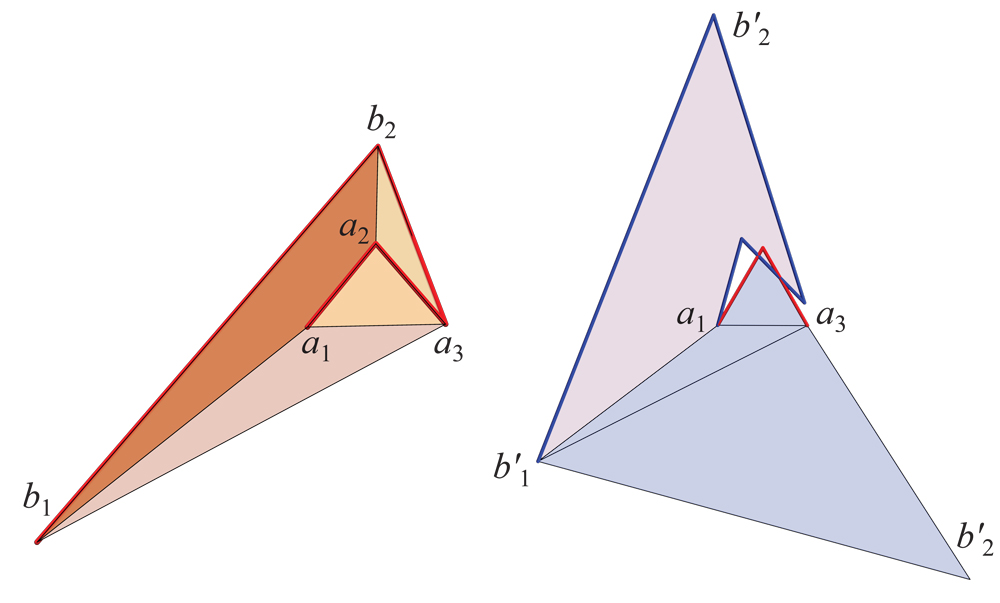}
\caption{(a)~$B$ and $\triangle$ share $a_3$. \texttt{Z} $= a_1 a_2 a_3 b_2$.
(b)~Unfolding with overlap.}
\figlab{PlatformNet_SmallF}
\end{figure}

This leaves the case where there are no two disjoint faces, nor two faces that share a single vertex:
every pair of faces shares two or more vertices.
If two faces share non-adjacent vertices, they cannot both be convex. So in fact the
condition is that each two faces share an edge.
Then, it is not difficult to see that $G$ can only be a tetrahedron,
as the following argument shows.

Suppose $B = b_1 b_2 \ldots b_k$ is $k$-gon.
Add one triangle $t_1 = a b_1 b_2$; see Fig.~\figref{SharingEdgesTetra}.
A second triangle must share an edge with $t_1$, say $b_2 a$,
so sharing with $B$ leads to $t_2 = a b_2 b_3$.
Now a third triangle must share with $B, t_1, t_2$. The only uncovered
edge of $t_1$ is $b_1 a$. But $t_3= a b_1 b_k$ does not share an edge with $t_2$
unless $k=3$.
In that case we have a tetrahedron.
\begin{figure}[htbp]
\centering
\includegraphics[width=0.5\linewidth]{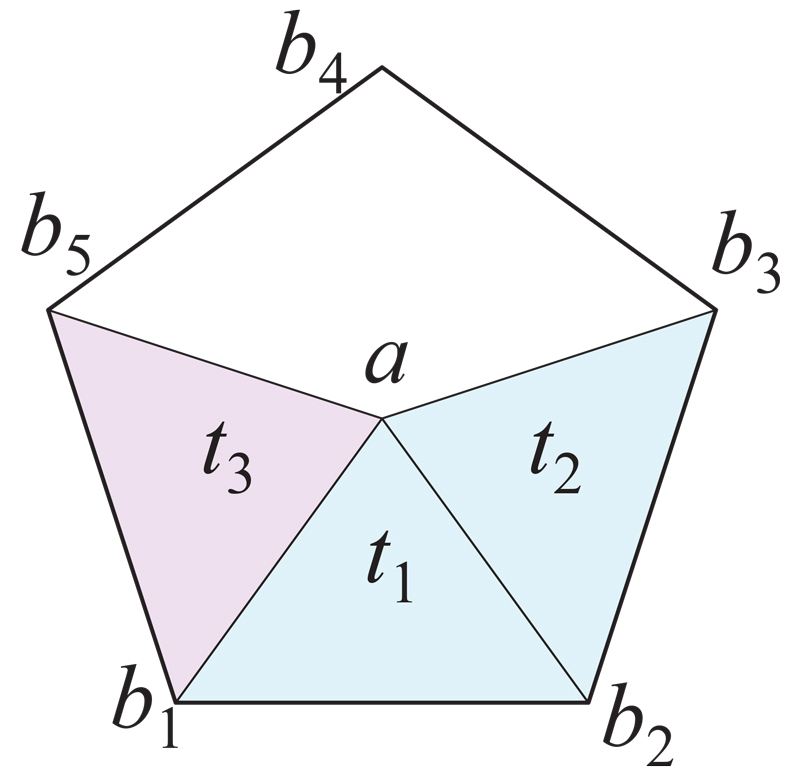}
\caption{Every pair of faces shares an edge.}
\figlab{SharingEdgesTetra}
\end{figure}

So the only case remaining is a tetrahedron. But it is well known that the thin, nearly flat
tetrahedron unfolds with overlap: Fig.~\figref{TetraOverlap}.
And since there is only one tetrahedron combinatorial type, this completes the inventory.
\begin{figure}[htbp]
\centering
\includegraphics[width=0.75\linewidth]{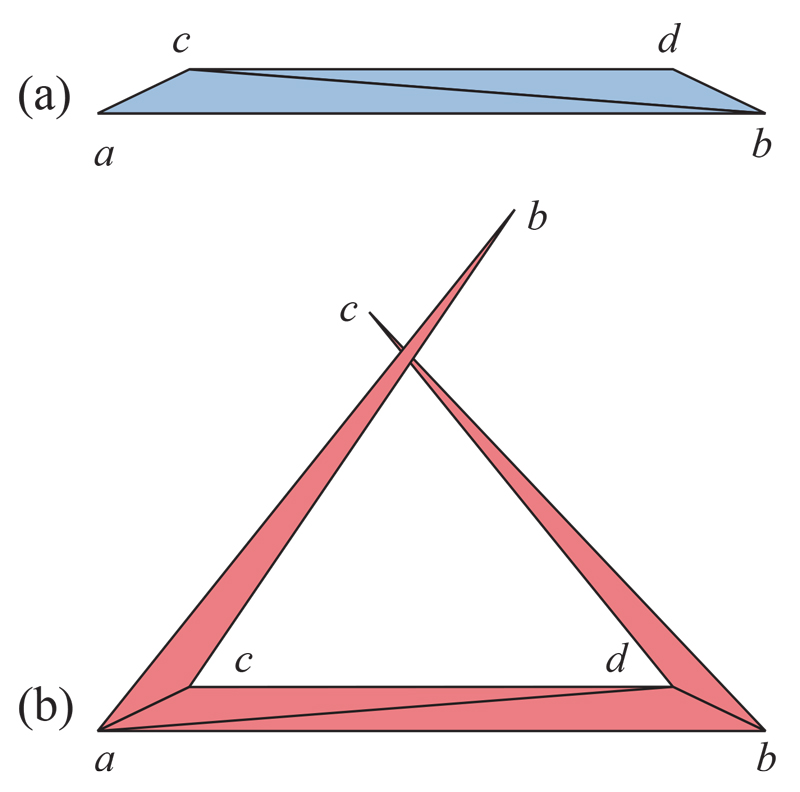}
\caption{Fig.~28.2 [detail], p.314 in~\protect\cite{do-gfalop-07}: tetrahedron overlap.
Blue: exterior. Red: interior.
Cut tree $T=abcd$.}
\figlab{TetraOverlap}
\end{figure}

\clearpage 

\section{Theorem}
\seclab{Theorem}
We have proved this theorem:

\begin{thm}
\thmlab{main}
Any $3$-connected planar graph $G$ can be realized as a convex polyhedron $P$
that has a spanning cut-tree $T$ such that the unfolding of $P \setminus T$
overlaps in the plane.
\end{thm}

So together with Ghomi's result,\footnote{
See~\cite{sert2018unfoldings} for a different proof of~\cite{g-aucp-14}.}
any combinatorial polyhedron type can be realized to
unfold and avoid overlap, or realized to unfold with overlap.

\medskip

Returning to Malkevitch's example of a combinatorial cube,
consider Fig.~\figref{CubeUnf}. Starting from the standard Schlegel diagram for a cube
(one square inside another ($B$), trapezoid faces between the squares),
horizontal scaling (step~(6) of Algorithm~\ref{Algo}) is applied to squeeze the top and bottom squares
to $1 \times 2$ and $2 \times 4$ diamonds, so that the angle at $a_2$ becomes small, 
in this case $2 \arctan(1/2) \approx 53^\circ$.
The lifting leaves the curvatures at $a_1,a_2$ to be small enough, $6.0^\circ, 6.5^\circ$,
so the vertical scaling step~(7) of Algorithm~\ref{Algo} is not needed.
\begin{figure}[htbp]
\centering
\includegraphics[width=0.85\linewidth]{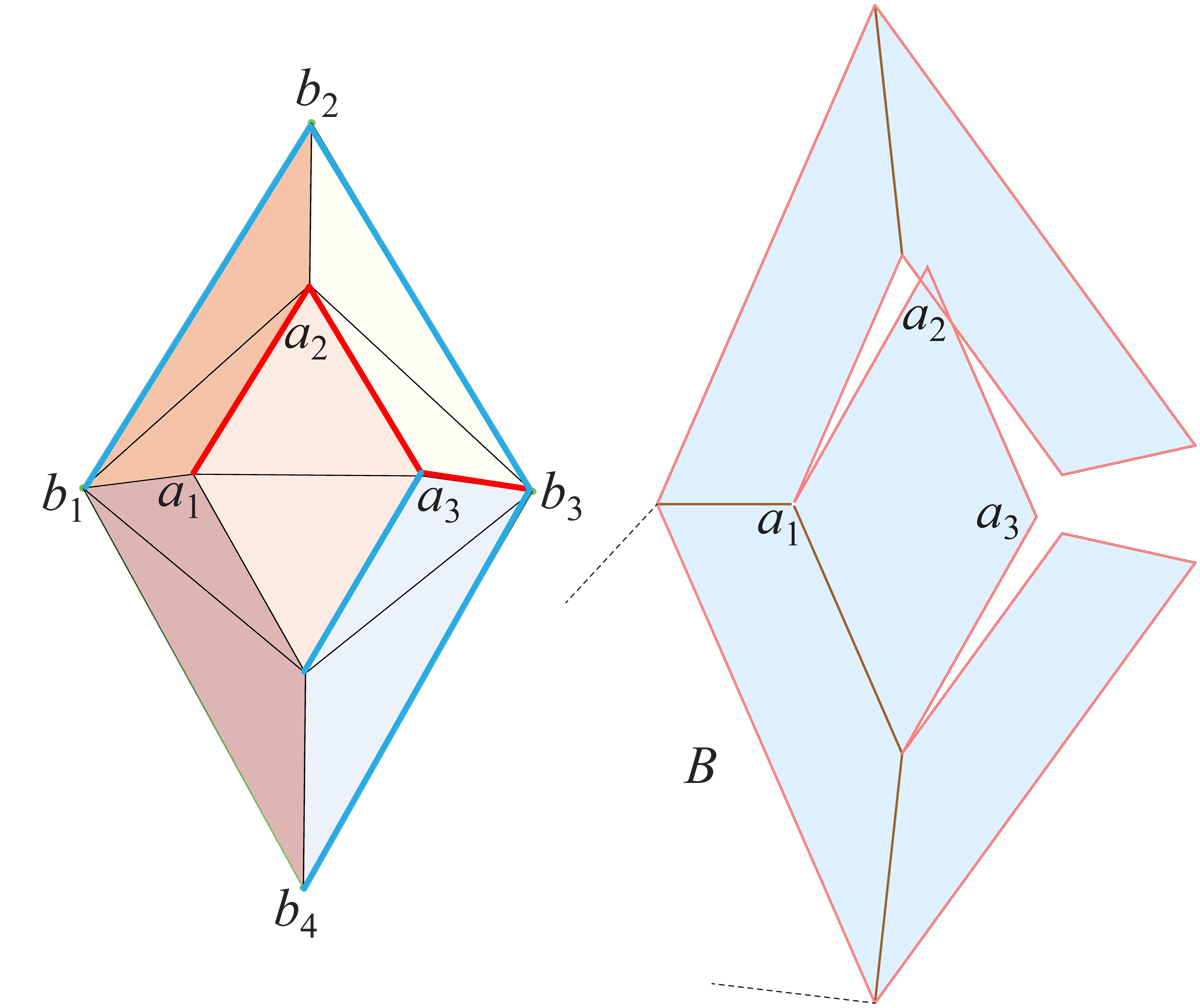}
\caption{Unfolding of a combinatorial cube. Diagonals in the left figure are an artifact of the software;
all faces are planar congruent trapezoids. Base $B$ attached left of $b_1 b_4$ not shown.
Vertex coordinates:
$$(-1, 0, 0.5), (1, 0, 0.5), (0, -2, 0.5), (0, 2, 0.5), (-2, 0, 0), (2, 0, 0), (0, -4, 0), (0, 4, 0)$$}
\figlab{CubeUnf}
\end{figure}

\paragraph{Acknowledgements.}
I benefitted from discussions with Richard Mabry and Joseph Malkevitch.


\bibliographystyle{alpha}
\bibliography{refs}
\end{document}